\documentclass[aps,prd,amsmath,amssymb,preprintnumbers,a4paper,11pt]{revtex4-1}
\pdfoutput=1
\usepackage{amsthm}
\usepackage{graphicx}
\usepackage[ margin=5pt, font=small,labelfont=bf, justification=raggedright]{caption}
\usepackage{url}
\usepackage[bookmarks, pagebackref=false]{hyperref}
\usepackage{color}
	\definecolor{rossoCP3}{cmyk}{0,.88,.77,.40}
		\definecolor{graa}{rgb}{0.8,0.8,0.8}
		\definecolor{blaa}{rgb}{0.2,0.2,0.6}
		\hypersetup{
			colorlinks,
			bookmarksopen,
			bookmarksnumbered,
			citecolor=blaa, 		
			linkcolor=rossoCP3,	
			urlcolor=rossoCP3,			
			}
\usepackage{dcolumn}
\usepackage{bm}
\usepackage{bbm}
\usepackage{pxfonts}

\usepackage{amsmath}
\usepackage{amsfonts}

\usepackage{dcolumn}
\usepackage{bm}
\usepackage{amssymb}
\usepackage{latexsym}
\usepackage{color}

\usepackage{epsfig}
\usepackage{placeins}

\usepackage{youngtab}
\usepackage{slashed}
\usepackage{subfig}

\begin{document}

\title{\Large  The unobservable axial-pion mixing \\ in QCD and near-conformal dynamics}
\author{Roshan Foadi}
\email{roshan.foadi@gmail.com}
\affiliation{Helsinki Institute of Physics, P.O. Box 64,\\
FI-000140, University of Helsinki, Finland}

\begin{abstract}
We study a chiral quark model of $\pi$, $\sigma$, $\rho$, and $a_1$ mesons in the large-$N_c$ limit. We show that the quadratic $a_1-\pi$ mixing can be set to zero at zero momentum, thus protecting the $g_{\sigma\pi\pi}$ and $g_{\rho\pi\pi}$ couplings from being contaminated by the corresponding vertices with one or both pions replaced by the axial-vector meson. We further require the chiral-quark Lagrangian to feature an approximate classical scale invariance, the latter being only broken by dimension-two mass terms, and that the longitudinal vector meson scattering amplitudes grow at most like $s$, where $\sqrt{s}$ is the center-of-mass energy. This allows us to accurately predict the correct value of $g_{\sigma\pi\pi}$, $g_{\rho\pi\pi}$, and the $a_1$ decay constant $f_{a_1}$. We further show that strongly-coupled theories with near-conformal dynamics are expected to feature an approximate custodial chiral symmetry, with interesting phenomenological consequences. 
\end{abstract}

\maketitle

\section{Introduction}
It has been known for many years that the effective Lagrangian of pions and and $\rho$ mesons accounts for the correct $g_{\rho\pi\pi}$ coupling \cite{Kaymakcalan:1984bz}, and guarantees unitarity of $\pi\pi$ scattering for energies below $m_\rho$. In \cite{Harada:1995dc} it was shown that a proper description of $\pi\pi$ scattering beyond $m_\rho$ requires introducing the scalar singlet $\sigma$, with a mass around 500 MeV. Recently, seeking for similarities between the $\sigma$ meson and the 125 GeV resonance observed at the LHC, it has been pointed out that the $\sigma\pi\pi$ coupling, in units of the pion decay constant, is equal to unity with remarkable accuracy \cite{Belyaev:2013ida}. In general, the physics of $\pi$ and $\sigma$ can be described in terms of a linear representation of the chiral group, with the $\pi^a$ isospin triplet and the $\sigma$ singlet forming a bi-doublet of $SU(2)_L\times SU(2)_R$. In fact the vertices with multiple $\pi$ and $\sigma$ legs can be made independent of each other by taking an arbitrarily large number of higher dimensional chiral invariants in the effective Lagrangian. However the experimental result $g_{\sigma\pi\pi}\simeq 1$ suggests that the higher dimensional operators are suppressed, and the linearly-realized effective Lagrangian of $\pi$ and $\sigma$ interactions is approximately invariant under classical scale invariance (CSI), the latter being only broken by a dimension-two mass term. In other words, experimental data suggest that the ordinary linear sigma model (LSM) is a good description of $\pi$ and $\sigma$ interactions. 

One could think of extending the particle content of the LSM to include the $\rho$ meson, and thus successfully account for the elastic $\pi\pi$ scattering below $m_\rho$ with a simple linear Lagrangian. This, however, does not work as smoothly as one could hope. The reason is that a linear $SU(2)_L\times SU(2)_R$ Lagrangian with vectors must also include the axial-vector meson $a_1$. If the the $a_1$ mass is assumed to be infinitely large, low-energy $\pi\pi$ data can be reproduced quite well within a linear model  \cite{Kaymakcalan:1984bz}. However, a finite $m_{a_1}$ introduces a non-zero $a_1-\pi$ quadratic mixing term, which can only be removed by shifting the axial-vector field. This modifies the $\rho\pi\pi$ and $\sigma\pi\pi$ vertices, and makes the agreement with experimental data possible but problematic: the $g_{\sigma\pi\pi}$ coupling turns out to equal to one by accident, rather than symmetry, and the longitudinal vector meson scattering amplitudes grow like $s^2$ rather than $s$.

As we shall see, a possible way to protect the $g_{\sigma\pi\pi}$ coupling from contamination by the $a_1-\pi$ mixing is to extended the ordinary custodial isospin symmetry to include a larger {\em custodial chiral symmetry} (CCS), which constrains the interactions between the spin-zero and spin-one resonances  \cite{Appelquist:1999dq}. This forbids the $a_1-\pi$ mixing, and, together with CSI, implies $g_{\sigma\pi\pi}=1$. Unfortunately, as we shall see, CCS also implies $g_{\rho\pi\pi}=0$, in clear contrast with experimental data.

In this paper we show that a linear model including spin-one resonances may account for the experimental data if chiral quark dynamics is added on top of the meson Lagrangian. This introduces running couplings in the Lagrangian of meson interactions, and allows setting the $a_1-\pi$ mixing to zero at zero momentum. With zero mixing for on-shell (or nearly on-shell) pions, the $g_{\sigma\pi\pi}$ and $g_{\rho\pi\pi}$ couplings receive no contribution from the corresponding vertices with one or both pions replaced by the axial-vector meson. As a consequence, and after imposing CSI, we obtain $g_{\sigma\pi\pi}=1$, while still having $g_{\rho\pi\pi}\neq 0$ at $\rho$ momenta greater than zero. We further require the longitudinal vector meson scattering amplitudes to grow at most like $s$, which allows us to impose an additional constraint on the parameters. As a result we are able to make accurate predictions for the axial decay constant $f_{a_1}$ and the $g_{\rho\pi\pi}$ coupling, which turn out to be in excellent agreement with data. Extrapolating these results to theories other than QCD, we argue that CCS is realized (approximately) in theories with near-conformal strongly-coupled dynamics, with interesting applications to dynamical electroweak symmetry breaking and LHC physics.

This paper is organized as it follows. In Sec. \ref{Sec:PionPion} we show how to extract the $g_{\sigma\pi\pi}$ and $g_{\rho\pi\pi}$ couplings from data, and present the most recent results. In Sec. \ref{Sec:Meson} we introduce the linearly-realized effective Lagrangian of $\pi$, $\sigma$, $\rho$ and $a_1$ mesons. We show that this may account for the experimental data, but does not explain why $g_{\sigma\pi\pi}\simeq 1$, and does not suppress the term growing like $s^2$ in longitudinal vector meson scattering amplitudes. In Sec. \ref{Sec:CQ} we extend the meson Lagrangian to include chiral quark dynamics, and compute, in the large-$N_c$ limit, the effect of quark dynamics on the meson interactions. Requiring the $a_1-\pi$ mixing to vanish for on-shell pions, and imposing that the  longitudinal vector meson scattering amplitudes grow at most like $s$, allows us to make accurate predictions for $f_{a_1}$ and $g_{\rho\pi\pi}$. In Sec. \ref{Sec:NC} we extrapolate our results to theories with near-conformal dynamics, and argue that these feature an approximate CCS. Finally in Sec. \ref{Sec:conclusions} we offer our conclusions.

\section{Pion coupling to the $\sigma$ and $\rho$ meson}\label{Sec:PionPion}
A good deal of knowledge in low-energy QCD is provided by the elastic $\pi\pi$ scattering. Well below 1 GeV this is unitarized by $\sigma$ and $\rho$ meson exchanges, resulting in the invariant amplitude
\begin{equation}
A(s,t,u)=-\left(\frac{1}{f_\pi^2}+\frac{4g_{\rho\pi\pi}^2}{m_\rho^2}\right)m_\pi^2
+\left(\frac{1}{f_\pi^2}-\frac{3g_{\rho\pi\pi}^2}{m_\rho^2}\right)s
-\frac{g_{\sigma\pi\pi}^2}{f_\pi^2}\frac{(s-2m_\pi^2)^2}{s-m_\sigma^2}
-g_{\rho\pi\pi}^2\left[
\frac{s-u}{t-m_\rho^2}+\frac{s-t}{u-m_\rho^2}
\right] \ ,
\label{Eq:A}
\end{equation}
where $f_\pi$ is the pion decay constant \cite{Rosner:2013ica}:
\begin{equation}
f_\pi= 92.2\pm 0.14\ {\rm MeV} \ .
\end{equation}
The $\sigma\pi\pi$ and $\rho\pi\pi$ couplings can be determined by extracting spin and isospin projections of $A(s,t,u)$ and comparing these with data. The isospin-zero and isospin-one projection of $A(s,t,u)$ are \cite{Bagger:1993zf}
\begin{eqnarray}
&& A_0(s,t,u)=3A(s,t,u)+A(t,s,u)+A(u,t,s)\ , \\
&& A_1(s,t,u)=A(t,s,u)-A(u,t,s)\ .
\end{eqnarray}
These can be further projected on spin-$J$ partial-waves by \cite{Sannino:1997fv}
\begin{equation}
A_{IJ}(s) = \frac{1}{64\pi}\sqrt{1-\frac{4m_\pi^2}{s}}\int_{-1}^1 d\cos\theta P_J(\cos\theta) A_I(s,t,u) \ .
\end{equation}
One can define a complex $\pi\pi$ coupling to a resonance as its pole residue,
\begin{equation}
g_{IJ}(s_{\rm pole})\equiv -16\pi\displaystyle{\lim_{s\to s_{\rm pole}}} (s-s_{\rm pole})A_{IJ}(s)\frac{2J+1}{(2p)^{2J}}\ ,
\end{equation}
where $s_{\rm pole}$ is complex. The $\sigma$ meson appears in the $I=0,\ J=0$ channel, whereas the $\rho$ exchanges contribute to the $I=1, J=1$ channel. Using (\ref{Eq:A}) allows us to relate $g_{\sigma\pi\pi}$ and $g_{\rho\pi\pi}$ respectively to $g_{00}(s_\sigma)$ and $g_{11}(s_\rho)$. The result is
\begin{eqnarray}
&& |g_{\sigma\pi\pi}| = \sqrt{\frac{2}{3}}\frac{|g_{00}(s_\sigma)|f_\pi}{|s_\sigma|}\left|1-\frac{4m_\pi^2}{s_\sigma}\right|^{-1/4}
\left|1-\frac{2m_\pi^2}{s_\sigma}\right|^{-1} \ , \label{Eq:gspp} \\
&& |g_{\rho\pi\pi}| = |g_{11}(s_\rho)|\left|1-\frac{4m_\pi^2}{s_\rho}\right|^{-1/4} \ . \label{Eq:grpp}
\end{eqnarray}
The values of $|g_{00}(s_\sigma)|$ and $ |g_{11}(s_\rho)|$ have been recently extracted from data in \cite{GarciaMartin:2011jx}. These, together with the values of $g_{\sigma\pi\pi}$ and $g_{\rho\pi\pi}$ obtained using (\ref{Eq:gspp}) and (\ref{Eq:grpp}), are shown in Tabs. \ref{Tab:sigmapipi} and \ref{Tab:rhopipi}. Note that $g_{\sigma\pi\pi}$ is very close to unity, as already observed in \cite{Belyaev:2013ida}. Here we assume that this is not an accident, but rather a direct consequence of an approximate CSI in the effective Lagrangian of meson interactions.

\begin{table}[t!]
\centering
\begin{tabular}{|c|c|c|c|}
\hline
Method & $\sqrt{s_\sigma}$ (MeV) & $g_{00}(s_\sigma)$ (GeV) & $\left|g_{\sigma\pi\pi}\right|$ \\
\hline
Roy & $445\pm 25-i\ 278^{+22}_{-18}$ & $3.4\pm 0.5$ & $1.0013\pm 0.17$ \\
\hline
GKPY & $457^{+14}_{-13}-i\ 279^{+11}_{-7}$ & $3.59^{+0.11}_{-0.13}$ & $1.0169\pm 0.06$ \\
\hline
\end{tabular}
\caption{Values of $s_\sigma$ and $g_{00}(s_\sigma)$ obtained in \cite{GarciaMartin:2011jx} using Roy and GKPY equations (see reference for details). The last column gives $g_{\sigma\pi\pi}$ as obtained from  (\ref{Eq:gspp}).}
\label{Tab:sigmapipi}
\vspace{0.5cm}
\begin{tabular}{|c|c|c|c|}
\hline
Method & $\sqrt{s_\rho}$ (MeV) & $g_{11}(s_\rho)$ & $\left|g_{\rho\pi\pi}\right|$ \\
\hline
Roy & $761^{+4}_{-3}-i\ 71.7^{+1.9}_{-2.3}$ & $5.95^{+0.12}_{-0.08}$ & $6.14 \pm 0.12$ \\
\hline
GKPY & $763^{+1.7}_{-1.5}-i\ 73.2^{+1.0}_{-1.1}$ & $6.01^{+0.04}_{-0.07}$ & $6.21\pm 0.07$ \\
\hline
\end{tabular}
\caption{Values of $s_\rho$ and $g_{11}(s_\rho)$ obtained in \cite{GarciaMartin:2011jx} using Roy and GKPY equations (see reference for details). The last column gives $g_{\rho\pi\pi}$ as obtained from  (\ref{Eq:grpp}).}
\label{Tab:rhopipi}
\end{table}
\section{Meson Lagrangian}\label{Sec:Meson}
We work in the context of two-flavor QCD, with an $SU(2)_L\times SU(2)_R$ chiral symmetry spontaneously broken to vectorial $SU(2)_V$ by the quark condensate. Consider first the lightest mesons, that is the $\pi^a$ triplet and the scalar singlet $\sigma$, and arrange them in an $SU(2)_L\times SU(2)_R$ spin-zero bi-doublet,
\begin{equation}
\Sigma\equiv\frac{1}{\sqrt2}\left(v+\sigma+i\ \pi^a\tau^a\right)\ ,
\label{Eq:Sigma}
\end{equation}
where $\tau^a$, $a=1,2,3$, are the Pauli matrices, and $v=f_\pi$ is the vacuum expectation value (VEV) breaking $SU(2)_L\times SU(2)_R$ to vectorial $SU(2)_V$.  Here and in remainder of this paper we shall neglect the pion mass, as this is expected to give a small contribution to the couplings. We can have vertices with arbitrary coefficients by taking an arbitrarily large number of invariants build out of the $\Sigma$ field and its derivatives. However we assume that the effective Lagrangian of $\pi$ and $\sigma$ interactions obeys CSI, the latter being only broken by a dimension-two mass term. This implies the LSM Lagrangian
\begin{equation}
L_{\rm LSM} = \frac{1}{2} {\rm Tr}\ \partial_\mu\Sigma\partial^\mu\Sigma^\dagger
+\frac{m^2}{2} {\rm Tr}\ \Sigma\Sigma^\dagger
-\frac{\lambda}{2}{\rm Tr}\ \Sigma\Sigma^\dagger\Sigma\Sigma^\dagger \ ,
\label{Eq:LSM}
\end{equation}
where $v^2=m^2/\lambda$. Using this to compute $A(s,t,u)$ leads to $g_{\sigma\pi\pi}=1$, in agreement with experiment. This result is a direct consequence of CSI: without the latter, higher dimensional operators could be added, and these would modify $g_{\sigma\pi\pi}$. The CSI hypothesis means that such operators are more suppressed than suggested by naive dimensional analysis.

\begin{figure}
\includegraphics[width=2.0in]{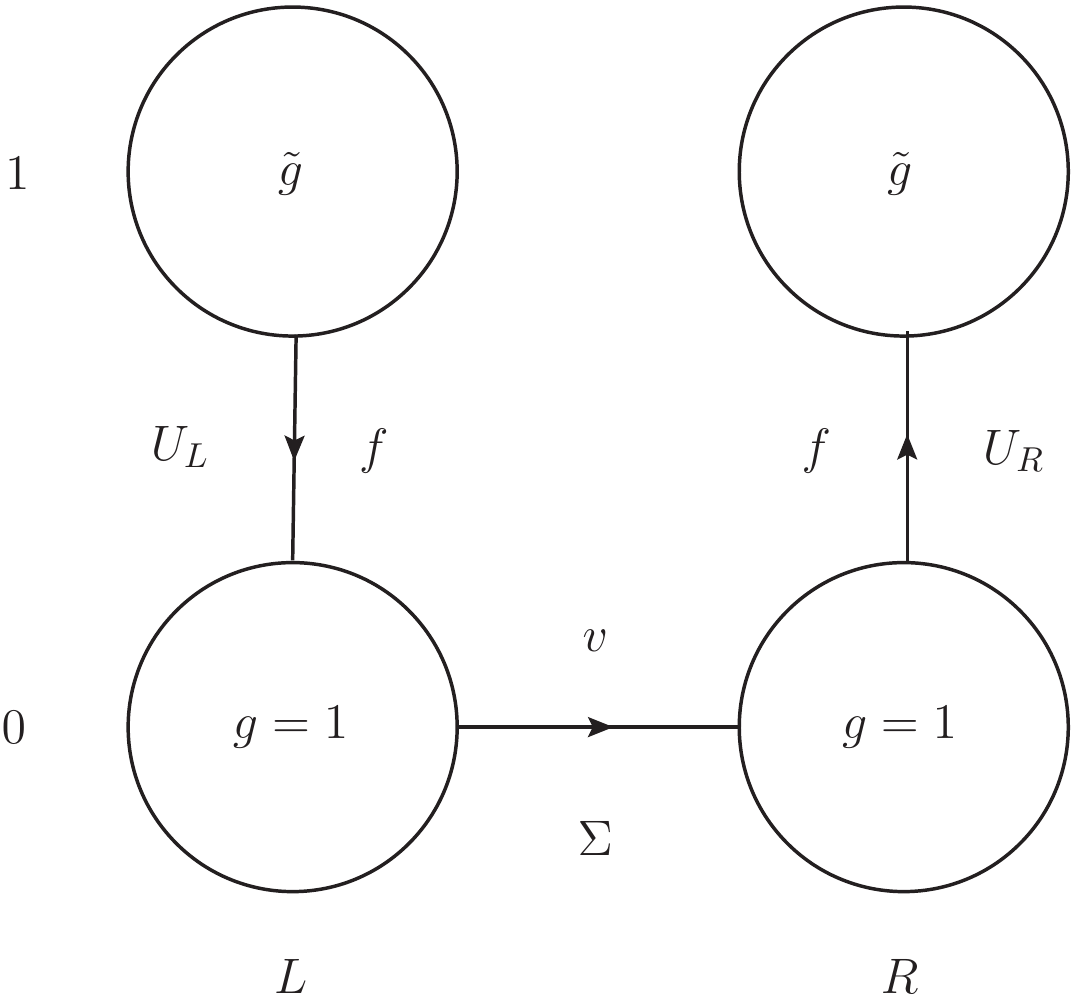}
\caption{Moose diagram for the meson Lagrangian of two-flavor QCD. The lower circles correspond to the ordinary $SU(2)_L\times SU(2)_R$ chiral symmetry, whereas the upper circles correspond to a gauged copy of the chiral symmetry group, whose gauge bosons are the $\rho$ and $a_1$ meson. The lower link corresponds to a linear sigma-model field containing the $\sigma$ meson and the pions. This field breaks the chiral symmetry to a vectorial $SU(2)_V$ subgroup. The left and right links correspond to non-linear sigma-model fields. These break the adjacent groups to a diagonal subgroup, and give mass to the spin-one mesons. See text for further details.}
\label{Fig:4Site}
\end{figure}
This simple scenario changes radically when the LSM is extended to include the lightest $SU(2)_L\times SU(2)_R$ spin-one resonances. These are conveniently, and without loss of generality, introduced as gauge bosons of a mirror chiral group. The full symmetry group becomes $SU(2)_{L0}\times SU(2)_{R0}\times SU(2)_{L1}\times SU(2)_{R1}$. This  is shown by the moose diagram of Fig.~\ref{Fig:4Site}, where circles represent $SU(2)$ groups and links represent  sigma fields breaking adjacent groups to a diagonal one. The ``1'' groups are the gauge groups of the spin-one fields $L_\mu^a$ and $R_\mu^a$, with gauge coupling $\tilde{g}$.  These fields can be expressed in terms of the vector and axial-vector mass eigenstates, that is the $\rho$ and $a_1$ meson, respectively. We shall call the corresponding fields $V_\mu^a$ and $A_\mu^a$:
\begin{equation}
L_\mu^a = \frac{V_\mu^a+A_\mu^a}{\sqrt2}\ , \quad
R_\mu^a = \frac{V_\mu^a-A_\mu^a}{\sqrt2} \ .
\end{equation}
The ``0'' groups correspond to the ordinary chiral symmetry. In order to extract chiral current correlators and decay constants, these groups have been gauged, with gauge fields ${\mathcal L}_\mu$ and ${\mathcal R}_\mu$, and gauge coupling $g=1$. The $\Sigma$ field is the same as in (\ref{Eq:Sigma}),  whereas $U_L$ and $U_R$ are nonlinear sigma fields with vacuum expectation value $f$. Note that the parity symmetry of the strong interactions forces  $\tilde{g}$ and $f$ to be equal for the ``left'' and ``right'' fields. The sigma-field covariant derivatives are
\begin{eqnarray}
D_\mu U_L &=& \partial_\mu U_L - i\ L_\mu\ U_L + i\ U_L\ {\mathcal L}_\mu\ , \nonumber \\
D_\mu U_R &=& \partial_\mu U_R  - i\  {\mathcal R}_\mu U_R + i\ U_R\ R_\mu\ , \nonumber \\
D_\mu \Sigma &=& \partial_\mu\Sigma - i\ {\mathcal L}_\mu\ \Sigma + i\ \Sigma\ {\mathcal R}_\mu\ ,
\end{eqnarray}
where $X_\mu\equiv X_\mu^a\tau^a/2$. The effective Lagrangian must be invariant under the parity transformations
\begin{equation}
L\leftrightarrow R\ , \quad 
{\mathcal L} \leftrightarrow {\mathcal R}\ , \quad 
U_L \leftrightarrow U_R^\dagger\ , \quad
\Sigma \leftrightarrow \Sigma^\dagger\  .
\end{equation}
Based on our assumptions, it must also contain mass and dimension-four terms only. This leads to the linearly independent invariants
\begin{eqnarray}
L_{\rm meson} &=& -\frac{1}{2\tilde{g}^2} {\rm Tr}\left(L_{\mu\nu}L^{\mu\nu}+R_{\mu\nu}R^{\mu\nu}\right)
+\frac{f^2}{4}{\rm Tr}\left(D_\mu U_L D^\mu U_L^\dagger + D_\mu U_R D^\mu U_R^\dagger\right) \nonumber \\
&-&\frac{i\ s_1}{\tilde{g}}{\rm Tr}\left[L_{\mu\nu}\left(D^\mu U_L D^\nu U_L^\dagger-D^\nu U_L D^\mu U_L^\dagger\right)
+R_{\mu\nu}\left(D^\mu U_R^\dagger D^\nu U_R-D^\nu U_R^\dagger D^\mu U_R\right)\right] \nonumber \\
&+&\frac{s_2}{2\tilde{g}^2}{\rm Tr}\left[\left(D_\mu U_L D^\mu U_L^\dagger\right)^2
+\left(D_\mu U_R D^\mu U_R^\dagger\right)^2\right]  \nonumber \\
&+&\frac{s_3}{2\tilde{g}^2}{\rm Tr}\left[ D_\mu U_L D_\nu U_L^\dagger D^\mu U_L D^\nu U_L^\dagger + D_\mu U_R^\dagger D_\nu U_R D^\mu U_R^\dagger D^\nu U_R \right]\nonumber \\
&+&\frac{1}{2} {\rm Tr}\ D_\mu\Sigma D^\mu\Sigma^\dagger
+\frac{m^2}{2} {\rm Tr}\ \Sigma\Sigma^\dagger
-\frac{\lambda}{2}{\rm Tr}\ \Sigma\Sigma^\dagger\Sigma\Sigma^\dagger \nonumber \\
&+&\frac{r_1}{2}{\rm Tr}\left(\Sigma\Sigma^\dagger D_\mu U_L^\dagger D^\mu U_L
+\Sigma^\dagger\Sigma D_\mu U_R D^\mu U_R^\dagger\right)
+r_2 {\rm Tr}\ D_\mu U_L^\dagger U_L \Sigma D^\mu U_R U_R^\dagger \Sigma^\dagger \nonumber \\
&-&\frac{r_3}{4}{\rm Tr}\left[D_\mu U_L^\dagger U_L
\left(\Sigma D^\mu\Sigma^\dagger-D^\mu\Sigma \Sigma^\dagger\right)
+D_\mu U_R U_R^\dagger\left(\Sigma^\dagger D^\mu\Sigma-D^\mu\Sigma^\dagger \Sigma\right)\right] \ ,
\label{Eq:Lmeson}
\end{eqnarray}
where $L_{\mu\nu}$ ($R_{\mu\nu}$) is the field-strength tensor for the gauge field $L_\mu$ ($R_\mu$). After symmetry breaking this Lagrangian is still invariant under a vectorial global symmetry $SU(2)_V$: this is the ordinary custodial isospin symmetry. In the limit $r_2,r_3\to 0$, there is an additional custodial symmetry, the $SU(2)_L^\prime\times SU(2)_R^\prime$ CCS mentioned in the introduction, with $L_\mu$ ($R_\mu$) transforming as a triplet of $SU(2)_L^\prime$ ($SU(2)_R^\prime$) \cite{Appelquist:1999dq}. Note that for global symmetries other than $SU(2)_L\times SU(2)_R$, CCS also requires the $r_1$ term to vanish: in the $SU(2)_L\times SU(2)_R$ case, the $r_1$ term does not violate CCS because $\Sigma\Sigma^\dagger$ is proportional to the identity matrix.

Now we express observables in terms of the Lagrangian parameters. The meson masses are
\begin{equation}
m_\sigma^2 = 2\lambda v^2\ , \quad
m_\rho^2 = \frac{\tilde{g}^2}{4}\left[f^2+(r_1-r_2)v^2\right]\ , \quad
m_{a_1}^2 = \frac{\tilde{g}^2}{4}\left[f^2+(r_1+r_2)v^2\right] \ .
\end{equation}
In order to compute the pion and axial-vector decay constant we first need to compute the axial-current correlator. This is found by inverting the axial-axial quadratic Lagrangian in momentum space,
\begin{equation}
L_A^{(2)}=\frac{1}{2}\left(\begin{array}{cc} {\mathcal A}_\mu^a & A_\mu^a\end{array}\right)
\left(D_A(q^2)\right)^{-1}
\left(\begin{array}{c} {\mathcal A}^{a\mu} \\ A^{a\mu}\end{array}\right) \ ,
\end{equation}
where ${\mathcal A}\equiv ({\mathcal L}-{\mathcal R})/\sqrt2$, and the inverse axial-matrix propagator $\left(D_A(q^2)\right)^{-1}$ can be easily extracted from (\ref{Eq:Lmeson}). The axial-axial correlator is
\begin{equation}
\Pi_{AA}(q^2)=\frac{1}{\left(D_A(q^2)\right)_{11}}\ .
\end{equation}
From this we can extract the pion and $a_1$ decay constants:
\begin{equation}
f_\pi^2=2\Pi_{AA}(0)\ , \quad
f_{a_1}^2=\lim_{q^2\to m_{a_1}^2} \left(1-\frac{m_{a_1}^2}{q^2}\right)\  2\Pi_{AA}(q^2)\ .
\end{equation}
Straightforward computation gives
\begin{equation}
f_\pi^2 = v^2\left(1-\frac{r_3^2\tilde{g}^2v^2}{8m_{a_1}^2}\right) \ , \quad
f_{a_1}^2 = \frac{2m_{a_1}^2}{\tilde{g}^2}\left(1-\frac{r_3\tilde{g}^2v^2}{4m_{a_1}^2}\right)^2 \ . \label{Eq:FA}
\end{equation}
The vector-current correlator is found by inverting the vector-vector quadratic Lagrangian in momentum space,
\begin{equation}
L_V^{(2)}=\frac{1}{2}\left(\begin{array}{cc} {\mathcal V}_\mu^a & V_\mu^a\end{array}\right)
\left(D_V(q^2)\right)^{-1}
\left(\begin{array}{c} {\mathcal V}^{a\mu} \\ V^{a\mu}\end{array}\right) \ ,
\end{equation}
where ${\mathcal V}\equiv ({\mathcal L}+{\mathcal R})/\sqrt2$. This gives
\begin{equation}
\Pi_{VV}(q^2)=\frac{1}{\left(D_V(q^2)\right)_{11}}\ .
\end{equation}
As the vector current is not broken, the $\Pi_{VV}$ correlator must vanish at $q^2=0$:
\begin{equation}
\Pi_{VV}(0)=0\ .
\label{Eq:PVV0}
\end{equation}
The $\rho$ decay constant is given by
\begin{equation} 
f_\rho^2=\lim_{q^2\to m_{\rho}^2} \left(1-\frac{m_{\rho}^2}{q^2}\right)\  2\Pi_{VV}(q^2)\ .
\end{equation}
Straightforward computation confirms (\ref{Eq:PVV0}), and gives
\begin{equation}
f_\rho^2 = \frac{2m_{\rho}^2}{\tilde{g}^2}\ .
\end{equation}

In order to evaluate $g_{\sigma\pi\pi}$ and $g_{\rho\pi\pi}$, we need to make sure that the quadratic terms are diagonal. However, from (\ref{Eq:Lmeson}) we obtain the $a_1-\pi$ mixing term
\begin{equation}
L_{A\pi}^{(2)} = -\frac{r_3\ v}{2\sqrt2}\ A_\mu^a \partial^\mu\pi^a \ .
\end{equation}
In order to remove this, we need to shift the axial-vector field by defining
\begin{equation}
A_\mu^a \equiv \tilde{A}_\mu^a + \frac{r_3\ v}{2\sqrt2\ m_{a_1}^2}\partial_\mu\pi^a \ ,
\end{equation}
where $\tilde{A}_\mu^a$ is the diagonal normalized field. Note that after the shift the pion field is non-canonically normalized. The normalized field $\tilde{\pi}$ is given by
\begin{equation}
\pi=\frac{v}{f_\pi}\tilde{\pi}^a\ .
\end{equation}

We can now evaluate $g_{\sigma\pi\pi}$ and $g_{\rho\pi\pi}$. The $\sigma\pi\pi$ vertices are
\begin{equation}
L_{\sigma\pi\pi} = -\frac{h_1\ m_\sigma}{2} \sigma \tilde{\pi}^a \tilde{\pi}^a
-\frac{h_2}{2m_\sigma} \sigma \partial_\mu \tilde{\pi}^a \partial^\mu \tilde{\pi}^a
+\frac{h_3}{m_\sigma} \partial_\mu \sigma \partial^\mu \tilde{\pi}^a \tilde{\pi}^a\ .
\end{equation}
Note that in the limit of zero $a_1-\pi$ mixing, $r_3\to 0$, only the first term survives. The other two terms arise respectively from the $\sigma a_1 a_1$ and $\sigma\pi a_1$ vertex. Computing the invariant amplitude $A(s,t,u)$ and extracting the $g_{\sigma\pi\pi}$ coupling gives \cite{Foadi:2008xj}
\begin{equation}
g_{\sigma\pi\pi} = \frac{f_\pi}{m_\sigma}\left(h_1-\frac{h_2}{2}-h_3\right)\ .
\end{equation}
From the Lagrangian we obtain
\begin{equation}
h_1 = \frac{m_\sigma v}{f_\pi^2}\ , \quad
h_2 = \frac{m_\sigma v}{f_\pi^2}\left(1-\frac{f_\pi^2}{v^2}\right)\left(1+\frac{m_\rho^2}{m_{a_1}^2}-\frac{r_1 \tilde{g}^2 v^2}{2 m_{a_1}^2}\right)\ , \quad
h_3 = \frac{m_\sigma v}{f_\pi^2}\left(1-\frac{f_\pi^2}{v^2}\right)\ ,
\end{equation}
whence
\begin{equation}
g_{\sigma\pi\pi}  = \frac{v}{f_\pi}\left[1-\left(1-\frac{f_\pi^2}{v^2}\right)\left(\frac{3}{2}+\frac{m_\rho^2}{2m_{a_1}^2}
-\frac{r_1 \tilde{g}^2 v^2}{4 m_{a_1}^2}\right)\right]\ .
\label{Eq:HppForm}
\end{equation}
Note that for arbitrary values of the parameters, $g_{\sigma\pi\pi} =1$ only in the limit $m_{a_1}\to\infty$.

The $\rho\pi\pi$ vertices are
\begin{equation}
L_{\rho\pi\pi} = g_1 \varepsilon^{abc} V_\mu^a \tilde{\pi}^b \partial^\mu \tilde{\pi}^c
+\frac{g_2}{m_{a_1}^2} \varepsilon^{abc} \partial_\mu V_\nu^a \partial^\mu\tilde{\pi}^b \partial^\nu\tilde{\pi}^c\ .
\end{equation}
On-shell, these give
\begin{equation}
g_{\rho\pi\pi} = g_1+\frac{m_\rho^2}{2m_{a_1}^2}g_2\ .
\end{equation}
The $g_1$ and $g_2$ couplings extracted from the Lagrangian are
\begin{equation}
g_1 = \frac{m_\rho^2 v}{m_{a_1} f_\pi^2}\sqrt{1-\frac{f_\pi^2}{v^2}}\ , \quad
g_2 = \frac{\tilde{g}(s_1-1)}{\sqrt2}\left(\frac{v^2}{f_\pi^2}-1\right) \ ,
\end{equation}
whence
\begin{equation}
g_{\rho\pi\pi} = \frac{m_\rho^2 v}{m_{a_1} f_\pi^2}\sqrt{1-\frac{f_\pi^2}{v^2}}
\left(1+\frac{(s_1-1)\tilde{g}v}{2\sqrt2 m_{a_1}} \sqrt{1-\frac{f_\pi^2}{v^2}} \right) \ .
\label{Eq:VppForm}
\end{equation}

We use $f_\pi$, $m_\sigma$, $m_\rho$, $m_{a_1}$, $f_\rho$ and $f_{a_1}$ as input, solve for $m$, $f$, $\tilde{g}$, $\lambda$, $r_2$ and $r_3$, and use the resulting expressions to evaluate $g_{\sigma\pi\pi}$ and $g_{\rho\pi\pi}$. We have already given the experimental value for $f_\pi$, $m_\sigma$ and $m_\rho$. For the $a_1$ mass we use the Particle Data Group estimate \cite{Beringer:1900zz}
\begin{equation}
m_{a_1} = 1230\pm 40\ {\rm MeV}\ .
\end{equation}
The $f_\rho$  and $f_{a_1}$ decay constants are measured from leptonic decays. From \cite{Bloch:1999vka} we obtain the values
\begin{eqnarray}
f_\rho &=& 152.5\pm 3.5 \ {\rm MeV}\ , \\
f_{a_1} &=& 143.5\pm 12.5\ {\rm MeV}\ . \label{Eq:fAExp}
\end{eqnarray}
For simplicity we average over the estimates for $m_\sigma$ and $m_\rho$ shown in Tabs. \ref{Tab:sigmapipi} and \ref{Tab:rhopipi}. Taking the central values to determine the Lagrangian parameters, and using the latter to evaluate $g_{\sigma\pi\pi}$ and $g_{\rho\pi\pi}$, we obtain
\begin{equation}
g_{\sigma\pi\pi}\simeq 0.09+0.13 r_1\ , \quad
g_{\rho\pi\pi} \simeq 4.5+1.2 s_1\ . \nonumber
\end{equation}
Agreement with experiment requires
\begin{equation}
r_1 \simeq 6.9\ ,\quad
s_1 \simeq 1.4\ . \nonumber
\end{equation}
However neither result is fully satisfactory. First, the $g_{\sigma\pi\pi}$ coupling appears to be equal to one by accident, rather than symmetry. Second, agreement with experiment for the $g_{\rho\pi\pi}$ coupling implies $s_1={\cal O}(1)$. This, in turn, spoils the special relationship between trilinear and quartic spin-one coupling leading to cancellation of the term growing like $s^2$, in longitudinal vector boson scattering amplitudes (where $\sqrt{s}$ is the center-of-mass energy). We would rather have $|s_1|\sim |s_2|\sim |s_3|\ll 1$, or else unitarity might be problematic already for $\sqrt{s}\gtrsim m_\rho$.  As an alternative approach, one can impose CCS. This implies $r_3=0$, whence $f_\pi=v$. Then we obtain $g_{\sigma\pi\pi}=1$, in agreement with experiment. Unfortunately this also gives $g_{\rho\pi\pi}=0$: in fact both $a_1-\pi$ mixing and $\rho\pi\pi$ vertex arise from the operator proportional to $r_3$. 

We end this section by noting that the linear sigma model with vector and axial-vector resonances has been recently analyzed in \cite{Parganlija:2012fy} for three-flavor QCD. There the chiral partner of the pions in the meson Lagrangian is taken to be the $f_0(1370)$ resonance, rather than the $\sigma$ meson. While this is reasonable, it nonetheless leaves open the question of why $g_{\sigma\pi\pi}\simeq 1$.
\section{Chiral-quark Lagrangian}\label{Sec:CQ}
Now we shall add chiral-quark dynamics on top of the meson Lagrangian (\ref{Eq:Lmeson}). We take the $q_L\equiv(u_L,d_L)$ and $q_R\equiv(u_R,d_R)$ doublets to respectively transform under $SU(2)_{L0}$ and $SU(2)_{R0}$, and impose that CSI is only broken by dimension-two mass terms.  Once again, the meaning of this is to assume that higher-dimensional operators are more suppressed than suggested by naive dimensional analysis. Imposing invariance under parity transformations the Lagrangian reads
\begin{eqnarray}
L_{\rm CQ} &=& \bar{q}_L i \slashed{D} q_L + \bar{q}_R i \slashed{D} q_R
-y\left(\bar{q}_L \Sigma q_R + \bar{q}_R \Sigma q_L\right)
+k\left(\bar{q}_L i \slashed{D}U_L^\dagger U_L q_L + \bar{q}_R i \slashed{D}U_R U_R^\dagger q_R\right) \nonumber \\
&-&\frac{1}{2\delta_{\tilde{g}}} {\rm Tr}\left(L_{\mu\nu}L^{\mu\nu}+R_{\mu\nu}R^{\mu\nu}\right)
+\frac{\delta_f}{4}{\rm Tr}\left(D_\mu U_L D^\mu U_L^\dagger + D_\mu U_R D^\mu U_R^\dagger\right) \nonumber \\
&+&\frac{\delta_Z}{2} {\rm Tr}\ D_\mu\Sigma D^\mu\Sigma^\dagger
+\frac{\delta_m}{2} {\rm Tr}\ \Sigma\Sigma^\dagger
-\frac{\delta_\lambda}{2}{\rm Tr}\ \Sigma\Sigma^\dagger\Sigma\Sigma^\dagger \nonumber \\
&+&\frac{\delta_{r_1}}{2}{\rm Tr}\left(\Sigma\Sigma^\dagger D_\mu U_L^\dagger D^\mu U_L
+\Sigma^\dagger\Sigma D_\mu U_R D^\mu U_R^\dagger\right)
+\delta_{r_2} {\rm Tr}\ D_\mu U_L^\dagger U_L \Sigma D^\mu U_R U_R^\dagger \Sigma^\dagger \nonumber \\
&-&\frac{\delta_{r_3}}{4}{\rm Tr}\left[D_\mu U_L^\dagger U_L
\left(\Sigma D^\mu\Sigma^\dagger-D^\mu\Sigma \Sigma^\dagger\right)
+D_\mu U_R U_R^\dagger\left(\Sigma^\dagger D^\mu\Sigma-D^\mu\Sigma^\dagger \Sigma\right)\right] \ ,
\label{Eq:CQ0}
\end{eqnarray}
where
\begin{equation}
\slashed{D} q_L = \left(\slashed{\partial}-i\ \slashed{\cal{L}}\right) q_L\ , \quad
\slashed{D} q_R = \left(\slashed{\partial}-i\ \slashed{\cal{R}}\right) q_R\ ,
\end{equation}
Now the $\pi$, $\sigma$, $\rho$ and $a_1$ mesons carry a substantial part of the strong force, and the terms contained in the meson Lagrangian parametrize residual gluon interactions below the chiral-symmetry-breaking scale.

We now evaluate the effect of chiral quark dynamics by computing the renormalized Lagrangian in the large-$N_c$ limit, with one fermion loop giving a contribution of the same order of the tree-level terms. For this model to be realistic not only at zero momentum, but also at the $\rho$ and $a_1$ mass scale, the fermion loops should be evaluated with a distribution-density of constituent quark masses: this insures confinement and avoids processes with unphysical multi-quark production \cite{Efimov:1993zg}. Here we shall follow a simpler approach. We evaluate the integrals in the unbroken phase using dimensional regularization, keeping the divergent terms only. We then make the replacement
\begin{equation}
\frac{1}{2-d/2} +\log 4\pi^2-\gamma \to \log\frac{\Lambda^2}{\mu^2}\ .
\end{equation}
Here $\Lambda$ is the cutoff, whose dependence must be removed by appropriately choosing the tree-level terms, whereas $\mu$ is the renormalization scale, which is of the order of the external momenta in the loop diagram. This approach allows us to compute the running parameters of the meson Lagrangian without having to deal with unphysical thresholds. 

Before evaluating the quark loops, we note that the operators proportional to $s_1$, $s_2$ and $s_3$ have been omitted in (\ref{Eq:CQ0}). As explained in the last section, these operators spoil the special relationship between trilinear and quartic vector meson couplings which lead to cancellation of the term growing like $s^2$, in longitudinal vector meson scattering amplitudes. In order to insure unitarity, we assume that these operators are zero in the large-$N_c$ limit. This in turn implies that $k=\pm 1$. In fact $k$ is the strength of the quark-vector interactions in units of the gauge coupling. Once the $\Lambda$-dependence in the vector-field kinetic terms is removed by appropriately choosing $\delta_{\tilde{g}}$, the $\Lambda$-dependence in the trilinear and quartic vertices is absorbed by the same counterterm,  in the fermion-bubble approximation, only if $k=\pm 1$. If instead $k\neq \pm1$, the $s_i$ terms must be introduced, and their running coefficients cannot be neglected to leading order in $N_c$.

\begin{figure}
\includegraphics[width=3.5in]{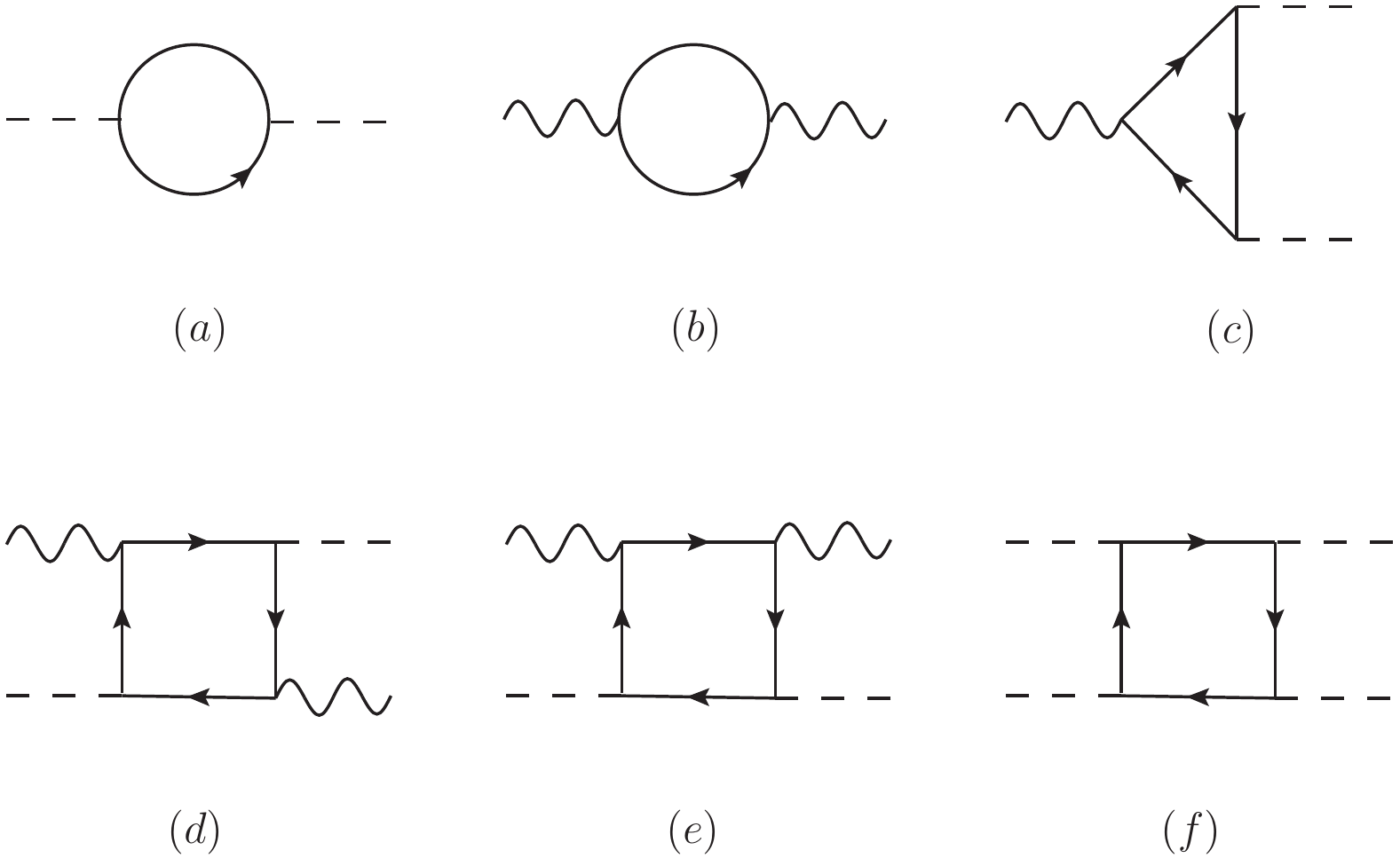}
\caption{Chiral quark contribution to the meson Lagrangian (\ref{Eq:CQ}), in the large-$N_c$ limit. The following diagrams contribute to the corresponding running parameters: (a) $Z(\mu^2)$, (b) $\tilde{g}^2(\mu^2)$, (c) $r_3(\mu^2)$, (d) $r_2(\mu^2)$, (e)  $r_1(\mu^2)$, (f) $\lambda(\mu^2)$. In dimensional regularization diagram (a) contributes with a momentum-independent $d=2$ pole, which is absorbed by the constant $m^2$ term. Because of gauge invariance, diagram (b) does not contribute to $f^2$, which is therefore just given by the residual gluonic term $\delta_f$ in (\ref{Eq:CQ0}).}
\label{Fig:Diagrams}
\end{figure}

The relevant quark loops are shown in Fig.~\ref{Fig:Diagrams} and lead to the running Lagrangian
\begin{eqnarray}
L_{\rm CQ}^\prime &=& \bar{q}_L i \slashed{D} q_L + \bar{q}_R i \slashed{D} q_R
-y\left(\bar{q}_L \Sigma q_R + \bar{q}_R \Sigma q_L\right)
+k\left(\bar{q}_L i \slashed{D}U_L^\dagger U_L q_L + \bar{q}_R i \slashed{D}U_R U_R^\dagger q_R\right) \nonumber \\
&-&\frac{1}{2\tilde{g}^2(\mu^2)} {\rm Tr}\left(L_{\mu\nu}L^{\mu\nu}+R_{\mu\nu}R^{\mu\nu}\right)
+\frac{f^2}{4}{\rm Tr}\left(D_\mu U_L D^\mu U_L^\dagger + D_\mu U_R D^\mu U_R^\dagger\right) \nonumber \\
&+&\frac{Z(\mu^2)}{2} {\rm Tr}\ D_\mu\Sigma D^\mu\Sigma^\dagger
+\frac{m^2}{2} {\rm Tr}\ \Sigma\Sigma^\dagger
-\frac{\lambda(\mu^2)}{2}{\rm Tr}\ \Sigma\Sigma^\dagger\Sigma\Sigma^\dagger \nonumber \\
&+&\frac{r_1(\mu^2)}{2}{\rm Tr}\left(\Sigma\Sigma^\dagger D_\mu U_L^\dagger D^\mu U_L
+\Sigma^\dagger\Sigma D_\mu U_R D^\mu U_R^\dagger\right)
+r_2(\mu^2) {\rm Tr}\ D_\mu U_L^\dagger U_L \Sigma D^\mu U_R U_R^\dagger \Sigma^\dagger \nonumber \\
&-&\frac{r_3(\mu^2)}{4}{\rm Tr}\left[D_\mu U_L^\dagger U_L
\left(\Sigma D^\mu\Sigma^\dagger-D^\mu\Sigma \Sigma^\dagger\right)
+D_\mu U_R U_R^\dagger\left(\Sigma^\dagger D^\mu\Sigma-D^\mu\Sigma^\dagger \Sigma\right)\right] \ ,
\label{Eq:CQ}
\end{eqnarray}
where
\begin{eqnarray}
&& Z(\mu^2)= \delta_Z+ \frac{N_c y^2}{8\pi^2}\log\frac{\Lambda^2}{\mu^2}\ , \quad
\frac{1}{\tilde{g}^2(\mu^2)} = \frac{1}{\delta_{\tilde{g}}}+ \frac{N_c k^2}{48\pi^2}\log\frac{\Lambda^2}{\mu^2}\ , \\
&& \lambda (\mu^2) = \delta_\lambda+ \frac{N_c y^4}{8\pi^2}\log\frac{\Lambda^2}{\mu^2}\ , \quad
m^2=\delta_m\ , \quad   f^2 = \delta_f\ ,  
\end{eqnarray}
and
\begin{eqnarray}
&& r_1(\mu^2) = \delta_{r_1} + \frac{N_c y^2 k^2}{8\pi^2}\log\frac{\Lambda^2}{\mu^2} \ ,  \\
&& r_2(\mu^2) = \delta_{r_2} + \frac{N_c y^2 k^2}{8\pi^2}\log\frac{\Lambda^2}{\mu^2} \ ,  \\
&& r_3(\mu^2) = \delta_{r_3} -\frac{N_c y^2 k}{4\pi^2}\log\frac{\Lambda^2}{\mu^2} \ .
\end{eqnarray}
The $\delta_f$ parameter is not renormalized by the quark loop because of gauge invariance, whereas $\delta_m$ absorbs a momentum-independent pole at $d=2$, but is otherwise left unchanged. We must choose the tree-level terms to remove the dependence on $\Lambda$. The $\delta_Z$ and $\delta_{\tilde{g}}$ parameters are fixed by imposing the {\em compositeness conditions}: the kinetic terms of the composite fields must vanish at a given {\em compositeness scale}. We allow for spin-zero and spin-one fields to have different compositeness scales, 
$\kappa_0$ and $\kappa_1$, respectively. This gives
\begin{eqnarray}
Z(\mu^2) =  \frac{N_c y^2}{8\pi^2}\log\frac{\kappa_0^2}{\mu^2}\ , \label{Eq:Z}\\
\frac{1}{\tilde{g}^2(\mu^2)} = \frac{N_c k^2}{48\pi^2}\log\frac{\kappa_1^2}{\mu^2}\ . \label{Eq:gtilde}
\end{eqnarray}
The $\delta_\lambda$ parameter is fixed by requiring the VEV to be independent of the scale $\mu$. Since the mass term is renormalized by $Z^{-1}(\mu^2)$ and the quartic coupling by $Z^{-2}(\mu^2)$, we require $\lambda(\mu^2)$ to approach zero like $\log\kappa_0^2/\mu^2$, as $\mu^2\to\kappa_0^2$:
\begin{eqnarray}
\lambda(\mu^2) = \frac{N_c y^4}{8\pi^2}\log\frac{\kappa_0^2}{\mu^2} = Z(\mu^2) y^2\ . \label{Eq:lambda}
\end{eqnarray}
The resulting VEV is
\begin{equation}
v^2=Z(\mu^2) \frac{m^2}{\lambda(\mu^2)}\ = \frac{m^2}{y^2}\ .
\end{equation}
The $\delta_{r_i}$ parameters can also be fixed by imposing renormalization conditions at the compositeness scale. This, however, leaves us with the ambiguity of which scale to use, $\kappa_0$ or $\kappa_1$. We find it much more convenient to renormalize the $r_i$ operators at zero momentum. The constituent quark mass, which is found by solving the Pagels-Stokar equation,
\begin{equation}
Z(m_q^2)m_q^2 = \frac{y^2 v^2}{2}\ ,
\end{equation}
acts as an infrared cutoff for the integrals. Therefore, we choose the $\delta_{r_i}$ parameters such that
\begin{eqnarray}
r_1(\mu^2) = r_1(m_q^2)-\frac{N_c y^2 k^2}{8\pi^2}\log\frac{\mu^2}{m_q^2} \ , \\
r_2(\mu^2) = r_2(m_q^2)-\frac{N_c y^2 k^2}{8\pi^2}\log\frac{\mu^2}{m_q^2} \ , \\
r_3(\mu^2) = r_3(m_q^2)+\frac{N_c y^2 k}{4\pi^2} \log\frac{\mu^2}{m_q^2}\ . \label{Eq:r3}
\end{eqnarray} 
Here we make a crucial ansatz for making the $a_1-\pi$ mixing unobservable. We require
\begin{equation}
r_3(m_q^2)=0\ . \label{Eq:nomix}
\end{equation}
In fact the $a_1-\pi$ mixing only affects the interaction vertices involving the pions. Since the latter are on-shell (for tree-level interactions), and (\ref{Eq:nomix}) implies that the $a_1-\pi$ mixing is zero for on-shell pions, it follows that the vertices are not affected. This implies that the $\sigma\pi\pi$ coupling receives no corrections from the $\sigma a_1 \pi$ and $\sigma a_1 a_1$ vertices, and is therefore forced by the hypothesis of CSI to be equal to one. On the other hand, the $\rho\pi\pi$ vertex is now nonzero, as $r_3(m_\rho^2)\neq 0$.  Furthermore, if $k>0$, $r_3(\mu^2)$ grows to positive values, and yields a negative contribution to the $f_{a_1}$ decay constant, as implied by (\ref{Eq:FA}). This is important, as the leading term is proportional to $m_{a_1}$, and alone would give an excessively large $f_{a_1}$. Since we argued that $|k|=1$, we must have
\begin{equation}
k=1\ .
\label{Eq:k}
\end{equation}

We now impose the constraints (\ref{Eq:Z})-(\ref{Eq:k}) and compute masses, decay constants, $g_{\sigma\pi\pi}$ and $g_{\rho\pi\pi}$. In doing so we should be careful to appropriately choose the value of $\mu^2$ in each case. In particular, for external momenta equal to $m_\sigma^2$ the fermion integrals, when evaluated in the broken phase, are dominated by the quark constituent mass, and thus we should take $\mu^2=m_q^2$ when computing $g_{\sigma\pi\pi}$ and $m_\sigma$. The latter is given by
\begin{equation}
m_\sigma^2 = \frac{2v^2\lambda(m_q^2)}{Z^2(m_q^2)} = 4 m_q^2 = \frac{16\pi^2}{N_c}\frac{v^2}{\log\kappa_0^2/m_q^2}\ .
\end{equation}
The $\rho$ and $a_1$ masses are, respectively,
\begin{eqnarray}
m_\rho^2 &=& \frac{\tilde{g}^2(m_\rho^2)}{4}\left[ f^2+\frac{r_1(m_\rho^2)-r_2(m_\rho^2)}{Z(m_q^2)}v^2\right] \nonumber \\
&=&\frac{12\pi^2}{N_c k^2 \log\kappa_1^2/m_\rho^2}\left[
f^2+\frac{v^2 k^2}{\log\kappa_0^2/m_q^2}\left(r_1(m_q^2)-r_2(m_q^2)\right)\right]\ ,
\label{Eq:rho}
\end{eqnarray}
and
\begin{eqnarray}
m_{a_1}^2 &=& \frac{\tilde{g}^2(m_{a_1}^2)}{4}\left[ f^2+\frac{r_1(m_{a_1}^2)+r_2(m_{a_1}^2)}{Z(m_q^2)}v^2\right] \nonumber \\
&=&\frac{12\pi^2}{N_c k^2 \log\kappa_1^2/m_{a_1}^2}\left[
f^2+\frac{v^2 k^2}{\log\kappa_0^2/m_q^2}\left(r_1(m_q^2)+r_2(m_q^2)-2\log\frac{m_{a_1}^2}{m_q^2}\right)\right] \ .
\label{Eq:a1}
\end{eqnarray}
Note that the $Z$ factor in these expressions is properly taken at $m_q^2$, since the $\rho\rho(v+\sigma)(v+\sigma)$ and $a_1 a_1(v+\sigma)(v+\sigma)$ interactions contributing to the mass are taken with the scalar legs at zero momentum.

The decay constants are
\begin{eqnarray}
f_\pi^2 &=& v^2 \ , \\
f_{a_1} ^2&=&  \frac{2m_{a_1}^2}{\tilde{g}^2(m_{a_1}^2)}\left[
1-\frac{\tilde{g}^2r_3(m_{a_1}^2)v^2}{4m_{a_1}^2 Z(m_q^2)}\right]^2 \nonumber  \\
&=& \frac{N_c k^2}{24\pi^2}m_{a_1}^2\log\frac{\kappa_1^2}{m_{a_1}^2}
\left[1-\frac{24\pi^2}{N_c k}\frac{v^2}{m_{a_1}^2}\frac{\log m_{a_1}^2/m_q^2}{\log\kappa_1^2/m_{a_1}^2 \log \kappa_0^2/m_q^2}\right]^2\ , \\
f_\rho^2 &=& \frac{2m_\rho^2}{\tilde{g}^2(m_\rho^2)} = \frac{N_c k^2}{24\pi^2}m_\rho^2\log\frac{\kappa_1^2}{m_\rho^2} \ .
\end{eqnarray}
Note that now, because of the constraint (\ref{Eq:nomix}), $f_\pi$ receives no correction from the $a_1-\pi$ mixing. 

CSI and the condition (\ref{Eq:nomix}) imply
\begin{equation}
g_{\sigma\pi\pi}=1\ ,
\end{equation}
in agreement with data. The $g_{\rho\pi\pi}$ coupling at $\rho$ momentum $\mu$ is
\begin{equation}
g_{\rho\pi\pi} = \frac{\tilde{g}(\mu^2) r_3(\mu^2)}{2\sqrt2 Z(m_q^2)}=\sqrt{\frac{24\pi^2}{N_c}}
\frac{\log\mu^2/m_q^2}{\sqrt{\log\kappa_1^2/\mu^2}\log\kappa_0^2/m_q^2}\ .
\end{equation}

\begin{figure}
\includegraphics[width=2.5in]{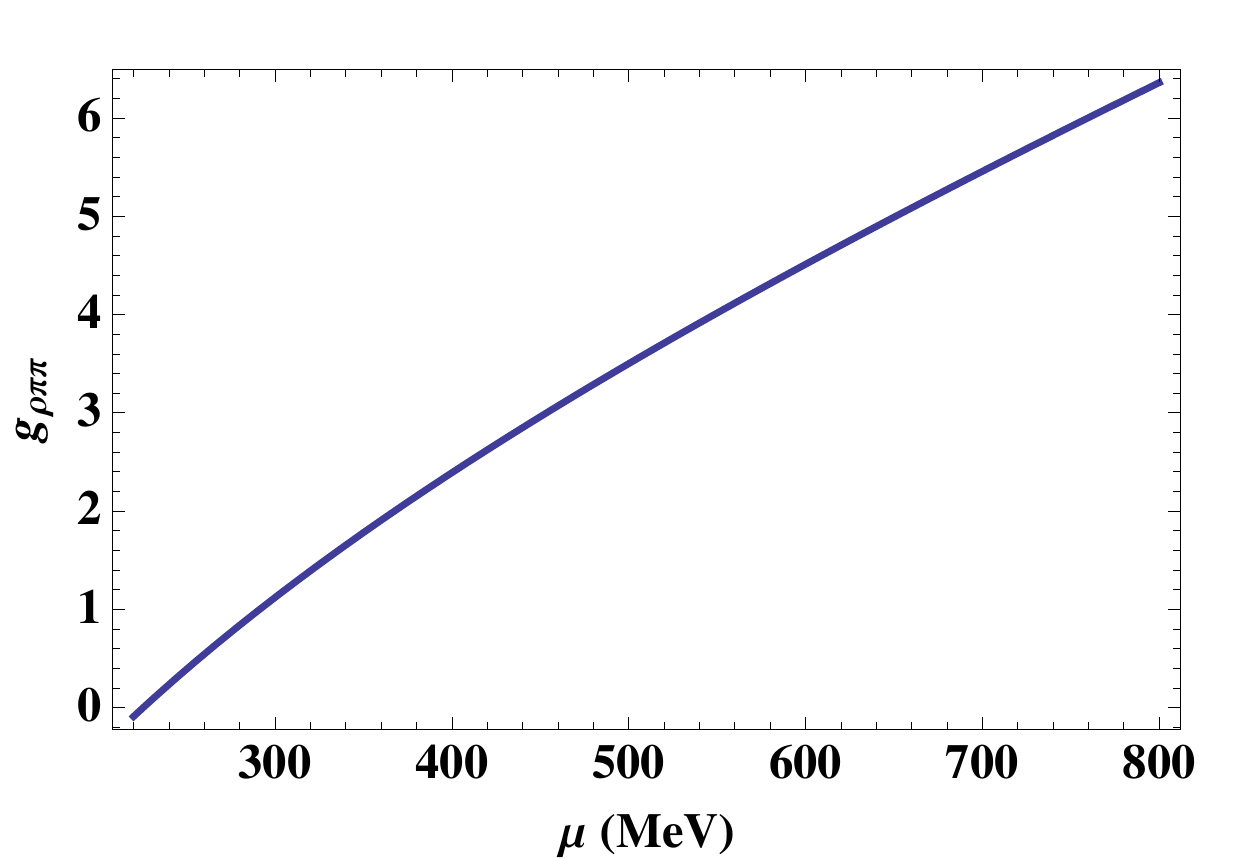}
\caption{The running $g_{\rho\pi\pi}$ coupling, as given by (\ref{Eq:grppRunning}). A zero momentum this is zero, because of the condition (\ref{Eq:nomix}), but grows to its correct experimental value at $\mu^2=m_\rho^2$.}
\label{Fig:grpp}
\end{figure}

We can now solve the equations for $f_\pi$, $m_\sigma$, and $f_\rho$ in terms of $m_q$, $\kappa_0$,  and $\kappa_1$, and substitute the results in the expressions for $f_{a_1}$ and $g_{\rho\pi\pi}$. This gives
\begin{eqnarray}
&& f_{a_1} = f_\rho\frac{m_{a_1}}{m_\rho}\xi(m_{a_1}^2)\left[1-\frac{f_{\pi}^2}{m_{a_1}^2}\frac{m_\rho^2}{f_\rho^2}\frac{\chi(m_{a_1}^2)}{\xi^2(m_{a_1}^2)}\right]\ , \\
&& g_{\rho\pi\pi}(\mu^2) = \frac{m_\rho}{f_\rho}\frac{\chi(\mu^2)}{\xi^2(\mu^2)}\ ,  \label{Eq:grppRunning}
\end{eqnarray}
where
\begin{equation}
\xi(\mu^2)\equiv \sqrt{1-\frac{N_c k^2}{24\pi^2}\frac{m_\rho^2}{f_\rho^2}\log\frac{\mu^2}{m_\rho^2}} \ , \quad
\chi(\mu^2)\equiv \frac{N_c k}{16\pi^2}\frac{m_\sigma^2}{f_\pi^2}\log\frac{4\mu^2}{m_\sigma^2}\ .
\end{equation}
Next we plug in the experimental values for $f_\pi$, $f_\rho$, $m_\rho$, $m_{a_1}$, and $m_\sigma$. For simplicity, we take an average of the values of $m_\sigma$ and $m_\rho$ shown in Tabs. \ref{Tab:sigmapipi} and \ref{Tab:rhopipi}, that is, the real part of $\sqrt{s_\sigma}$ and $\sqrt{s_{\rho}}$, respectively. Using (\ref{Eq:k}) this gives
\begin{eqnarray}
&& f_{a_1} = 142.5 \pm 10.5\ {\rm MeV} \ , \\
&& g_{\rho\pi\pi} = 5.52 \pm 0.31 \ ,
\end{eqnarray}
where $g_{\rho\pi\pi}$ has been evaluated at $\mu^2=m_\rho^2$. The prediction for $f_{a_1}$ agrees with the experimental value (\ref{Eq:fAExp}) at the 1$\sigma$ level, whereas $g_{\rho\pi\pi}$ agrees with the values shown in Tab. \ref{Tab:rhopipi} at the 2$\sigma$ level. In order to obtain an even more accurate result, we should take into account, in the effective Lagrangian, the pion mass and other sources of explicit chiral symmetry breaking. This, however, would force us to fit, rather than predict. Alternatively, we can assume that small departures from the chiral limit, in two-flavor QCD, affect the pion mass and decay constant, but nearly leaves the other observables unaffected. In this case we can use as input the value of $f_\pi$ extrapolated to the chiral limit, $f_\pi^0=88.3\pm 1.1$ MeV \cite{Gerber:1988tt}. The $f_{a_1}$ decay constant is unaffected, whereas for $g_{\rho\pi\pi}$ this gives
\begin{equation}
g_{\rho\pi\pi} = 6.02 \pm 0.37 \ ,
\end{equation}
in excellent agreement with the experimental value. This result for $g_{\rho\pi\pi}$ is especially interesting, since the coupling grows from zero at zero momentum, because of the condition (\ref{Eq:nomix}), to its correct experimental value at $\mu^2=m_\rho^2$, as shown in Fig.~\ref{Fig:grpp}. This and the result for $f_{a_1}$ suggest that the ansatz (\ref{Eq:nomix}) is a correct one. We finally notice that $f_{a_1}$ and $g_{\rho\pi\pi}$ are rather sensitive to $k$. For instance, setting $k=0.9$ ($k=1.1$) leads to a central value $f_{a_1}\simeq 159$ MeV ($f_{a_1}\simeq 123$ MeV). Therefore, the $k=1$ constraint appears to be satisfied to a good degree of accuracy. This implies that the $\rho$ and $a_1$ meson behave like true gauge bosons.
\section{Near-conformal dynamics}\label{Sec:NC}
In a strongly-coupled theory with near-conformal dynamics, the vector and axial-vector mesons corresponding to $\rho$ and $a_1$, respectively, are expected to be near-degenerate: $M_V\simeq M_A$ \cite{Kurachi:2006ej}. From (\ref{Eq:rho}) and (\ref{Eq:a1}) we obtain
\begin{equation}
M_A^2\log\frac{K_1^2}{M_A^2}-M_V^2\log\frac{K_1^2}{M_V^2} = \frac{24\pi^2 V^2}{N \log K_0^2/M_Q^2} r_2(M_A^2)\ .
\label{Eq:MAMV}
\end{equation}
Here $N$, $M_Q$, $V$, $K_0$ and $K_1$ are quantities in all equivalent to their QCD counterparts, respectively $N_c$, $m_q$, $v$, $\kappa_0$, and $\kappa_1$. The vector-axial near-degeneracy implies
\begin{equation}
r_2(M_A^2)\simeq 0 \ .
\end{equation}
Near-conformal dynamics also suggests that the $r_i$ functions run slowly. Since $r_2$ is small at scales near $M_A$, we expect it to be small for all external momenta. Furthermore, we still expect $r_3$ to vanish at zero momentum, that is, for $\mu=M_Q$: as we have seen this appears to be the proper way to remove the unwanted $A^a-\Pi^a$ mixing, where $\Pi^a$ are the pseudoscalars equivalent to the QCD pions.  Near-conformal dynamics suggests that $r_3(\mu^2)\simeq 0$ for all values of $\mu$ between $M_Q$ and the compositeness scale. Since $r_2(\mu^2)\simeq r_3(\mu^2)\simeq 0$, we see that CCS is expected to be an approximate symmetry of strongly-coupled theories with near-conformal dynamics. This has important physical consequences. For instance, we immediately obtain that, unlike QCD,
\begin{equation}
g_{V\Pi\Pi}\simeq 0\ .
\label{Eq:grppWalking}
\end{equation}
For the $H\Pi\Pi$ coupling, where $H$ is the scalar singlet equivalent to the $\sigma$ meson, we still obtain
\begin{equation}
g_{H\Pi\Pi}=1\ ,
\label{Eq:HPP}
\end{equation}
as this is a consequence of CSI in the meson Lagrangian. In technicolor theories $H$ is identified with the composite Higgs boson, whereas $\Pi^\pm$ and $\Pi^0$ are, respectively, the Goldstone boson eaten by the longitudinal component of the $W$ and $Z$ boson. Equation  (\ref{Eq:HPP}) suggests that the composite Higgs has standard coupling to the $W$ and $Z$ boson \cite{Belyaev:2013ida}, in agreement with observation. This is particularly interesting, since near-conformal technicolor theories are expected to feature a lighter scalar singlet than QCD-like technicolor. Including the large and negative radiative corrections to its mass from the top quark \cite{Foadi:2012bb}, such a state could be as light as the observed 125 GeV resonance. If the technicolor scenario is realized in Nature, (\ref{Eq:grppWalking}) tells us that the spin-one resonances might be difficult to find at the LHC, since the typically dominant di-boson decays are suppressed.
\section{Conclusions}\label{Sec:conclusions}
In this note we have discussed the possibility that the QCD $\pi^a$ pseudoscalar triplet and the $\sigma$ singlet form a linear bi-doublet of the chiral group $SU(2)_L\times SU(2)_R$. This is motivated by the fact that the $g_{\sigma\pi\pi}$ coupling, as implicitly defined in (\ref{Eq:A}), is, to a very good approximation, equal to one. While a LSM Lagrangian with pions and $\sigma$ does lead to $g_{\sigma\pi\pi}=1$ (neglecting small corrections from explicit chiral symmetry breaking), the situation is very different when the lightest $SU(2)_L\times SU(2)_R$ resonances are included. The quadratic mixing between pion and axial-vector meson requires a shift of the latter. This, in turn, introduces contamination of the $\sigma\pi\pi$ vertex from $\sigma a_1 a_1$ and $\sigma \pi a_1$, and leads, in general, to $g_{\sigma\pi\pi}\neq 1$. Furthermore, agreement with data for the $g_{\rho\pi\pi}$ coupling requires the inclusion of $O(p^4)$ operators other than the ones containing the kinetic terms. This spoils the special relationship between trilinear and quartic vector couplings leading to cancellation of the term growing like $s^2$, in longitudinal vector meson scattering amplitudes, and is therefore potentially problematic for unitarity. Setting the axial-pion mixing to zero does lead to $g_{\sigma\pi\pi}=1$, but also $g_{\rho\pi\pi}=0$, in clear contrast with experimental data.

The scenario changes radically when chiral quark dynamics is added on top of the meson Lagrangian. Working in the large-$N_c$ approximation, we have shown that fermion loops introduce momentum dependence on the couplings, and allows setting the axial-pion mixing to zero at zero momentum only. As a consequence the $\sigma\pi\pi$ vertex receives no more contribution from $\sigma a_1 a_1$ and $\sigma \pi a_1$. Then, allowing for classical scale invariance to be only broken by dimension-two mass terms implies $g_{\sigma\pi\pi}=1$. Furthermore, requiring the longitudinal vector meson scattering amplitudes to grow at most like $s$ leads to sensible predictions for $g_{\rho\pi\pi}$ and the axial-vector decay constant $f_{a_1}$, which turn out to be in excellent agreement with experimental data.

We finally argued that the vanishing of the axial-pion mixing at zero momentum suggests that meson Lagrangians of strongly-coupled theories with near-conformal dynamics are expected to feature an approximate chiral custodial symmetry, in addition to the ordinary custodial isospin symmetry. This has interesting phenomenological consequences when applied to near-conformal technicolor, such as, for instance, a small coupling of the spin-one resonances to the weak bosons. This, in turn, implies that the search for spin-one resonances at the LHC might be harder than expected.
\section*{Acknowledgements}
I would like to thank M. T. Frandsen, F. Sannino and K. Tuominen for advice and useful discussions.

\end{document}